\documentclass[amsmath,aps,prb]{revtex4}
\usepackage{epsfig}
\usepackage{pstricks}

\begin{document}

\title{Directed Sub-Wavelength Imaging Using a Layered Metal-Dielectric System}
\author{Wood, B.}
\author{Pendry, J. B.}
\affiliation{Blackett Laboratory, Imperial College, Prince Consort Road, London
SW7 2BW, United Kingdom}
\author{Tsai, D. P.}
\affiliation{Department of Physics, National Taiwan University, Taipei, Taiwan 
10617, Republic of China}

\date{\today}

\begin{abstract}
We examine some of the optical properties of a metamaterial consisting of 
thin layers of alternating metal and dielectric. We can model this material as 
a homogeneous effective medium with anisotropic dielectric permittivity. When
the components of this permittivity have different signs, the behavior of the
system becomes very interesting: the normally evanescent parts of a P-polarized
incident field are now transmitted, and there is a preferred direction of
propagation.

We show that a slab of this material can form an image with sub-wavelength 
details, at a position which depends on the frequency of light used. 
The quality of 
the image is affected by absorption and by the finite width of the layers; we 
go beyond the effective medium approximation to predict how thin the layers 
need to be in order to obtain subwavelength resolution.

\end{abstract}

\maketitle

\section{Introduction}
An anisotropic material in which one of the components of the
dielectric permittivity tensor has a different sign to the others 
has interesting properties. It 
supports the propagation of modes which would normally be evanescent, and 
these modes travel in a preferred direction. The propagation of 
evanescent modes gives us hope that an image produced by light travelling 
through a slab of such a material might retain a sharp profile; also, because 
the preferred
direction depends on the ratio of the components of the permittivity tensor,
it can be controlled by varying the frequency of light used.

We first look at a way of producing a metamaterial with the desired
properties: by making a system of thin, alternating metal and dielectric layers.
A system  of this type was proposed by \citet{ramakrishna:layered_lens} 
as a form of ``superlens"; it improves on the original suggestion for
a superlens,\citep{pendry:perfect_lens} which consists of just a single layer 
of metal, and has recently been realised.\cite{zhang:silver_superlens,
blaikie:super_resolution_silver} 

We then look at the dispersion relation for our anisotropic material, to  
see why modes which would be evanescent in both the metal and the dielectric 
separately are able to propagate in the combined system, and why there is a 
preferred direction of propagation.
The subwavelength details of the source are transmitted through the system 
because they couple to the surface plasmons 
\citep{ritchie:plasma_losses_thin_films} that exist on the boundaries between
metal and dielectric; this mechanism is the basis for the current interest in 
metallic structures for super-resolution imaging at optical frequencies.
\citep{pendry:perfect_lens, shalaev:composite_lens,
kawata:nanorod_array, zhang:silver_superlens, blaikie:super_resolution_silver}

Next, we investigate the transmission properties of a slab of this material,
and apply our formulae to the case of a line source. We show that
we can expect to obtain a sharp image as long as the amount of
absorption is not too high.

Finally, we go beyond the effective medium approximation to show the effect
of the finite layer widths on the optical properties. We demonstrate that the 
``resolution'' of the slab is limited by the width of the layers; thinner 
sheets mean that the description of the system using the effective medium
becomes increasingly accurate, and the image quality improves.

\section{Layered systems}
\begin{figure}
\begin{center}
\begin{pspicture}(0,0)(12,6)
\newrgbcolor{lightgrey}{0.8 0.8 0.8}
\newrgbcolor{vlightgrey}{0.9 0.9 0.9}
\psframe[linecolor=vlightgrey,fillstyle=solid,fillcolor=vlightgrey](1,0)(2.5,5)
\psframe[linecolor=lightgrey,fillstyle=solid,fillcolor=lightgrey](2.5,0)(3.5,5)
\psframe[linecolor=vlightgrey,fillstyle=solid,fillcolor=vlightgrey](3.5,0)(5,5)
\psframe[linecolor=lightgrey,fillstyle=solid,fillcolor=lightgrey](5,0)(6,5)
\psframe[linecolor=vlightgrey,fillstyle=solid,fillcolor=vlightgrey](6,0)(7.5,5)
\psframe[linecolor=lightgrey,fillstyle=solid,fillcolor=lightgrey](7.5,0)(8.5,5)
\psframe[linecolor=vlightgrey,fillstyle=solid,fillcolor=vlightgrey](8.5,0)(10,5)
\psframe[linecolor=lightgrey,fillstyle=solid,fillcolor=lightgrey](10,0)(11,5)
\psline[linecolor=black,linewidth=0.6mm]{->}(1,0)(2,0)
\rput(2,-0.2){\psframebox[linestyle=none,fillstyle=none]{$z$}}
\psline[linecolor=black,linewidth=0.6mm]{->}(1,0)(1,1)
\rput(0.8,1){\psframebox[linestyle=none,fillstyle=none]{$x$}}
\psline[linecolor=black,linewidth=0.6mm]{<->}(1,5.5)(2.5,5.5)
\rput(1.75,5.8){\psframebox[linestyle=none,fillstyle=none]{$d_1$}}
\psline[linecolor=black,linewidth=0.6mm]{<->}(2.5,5.5)(3.5,5.5)
\rput(3.0,5.8){\psframebox[linestyle=none,fillstyle=none]{$d_2$}}
\rput(5.5,3){\psframebox[linestyle=none,fillstyle=none]{$\epsilon_2$, $\mu_2$}}
\rput(6.75,2){\psframebox[linestyle=none,fillstyle=none]{$\epsilon_1$, $\mu_1$}}
\end{pspicture}
\end{center}
\caption{System geometry. The layers are infinite in extent in the $xy$-plane.}
\label{fig:layered-system}
\end{figure}
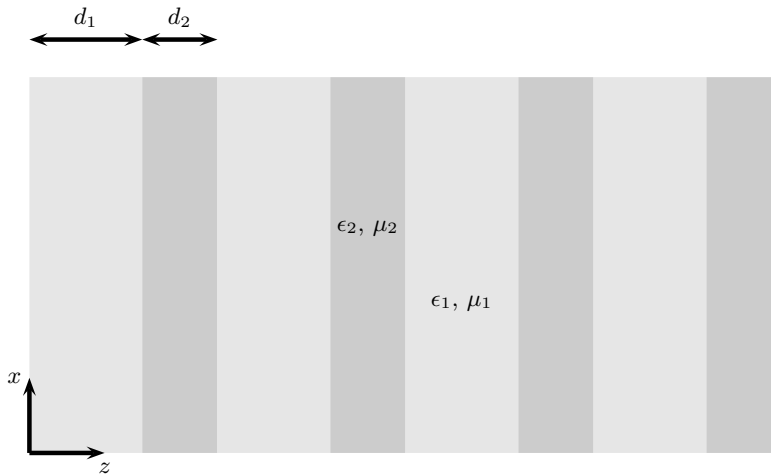

We concentrate on periodic layered systems of the form shown in figure 
\ref{fig:layered-system}.
We assume that each layer can be described by homogeneous and isotropic 
permittivity and permeability parameters. 
When the layers are sufficiently thin, we can treat the whole
system as a single anisotropic medium with the dielectric
permittivity\citep{rytov:em_stratified_medium, bergman:dielectric_composite}
\begin{align}
\epsilon_{x}&=\epsilon_y=\frac{\epsilon_1+\eta\epsilon_2}{1+\eta}
\label{eq:epsx}\\
\frac{1}{\epsilon_z}&=\frac{1}{1+\eta}\left(\frac{1}{\epsilon_1}+\frac{\eta}{
\epsilon_2}\right),
\label{eq:epsz}
\end{align}
where $\eta$ is the ratio of the two layer widths:
\begin{equation}
\eta=\frac{d_2}{d_1} .
\end{equation}
A helpful way to see this is through the characteristic matrix formalism;
\citep{born:principles} this method, which is related to that used by Rytov
in the original derivation,\citep{rytov:em_stratified_medium} 
is described in the appendix.

The homogenized magnetic permeability is given by expressions 
analogous to \eqref{eq:epsx} and \eqref{eq:epsz}. When $\eta$ is small, the 
effective parameters are dominated by the first medium, while for large $\eta$,
they resemble those of the second medium.

Only the ratio of the thicknesses of the two layers appears in the 
homogenized version, not the absolute value; however, the characterization of
the material using the effective medium parameters is more accurate when both
$d_1$ and $d_2$ are small. 

\begin{figure}
\epsfig{file=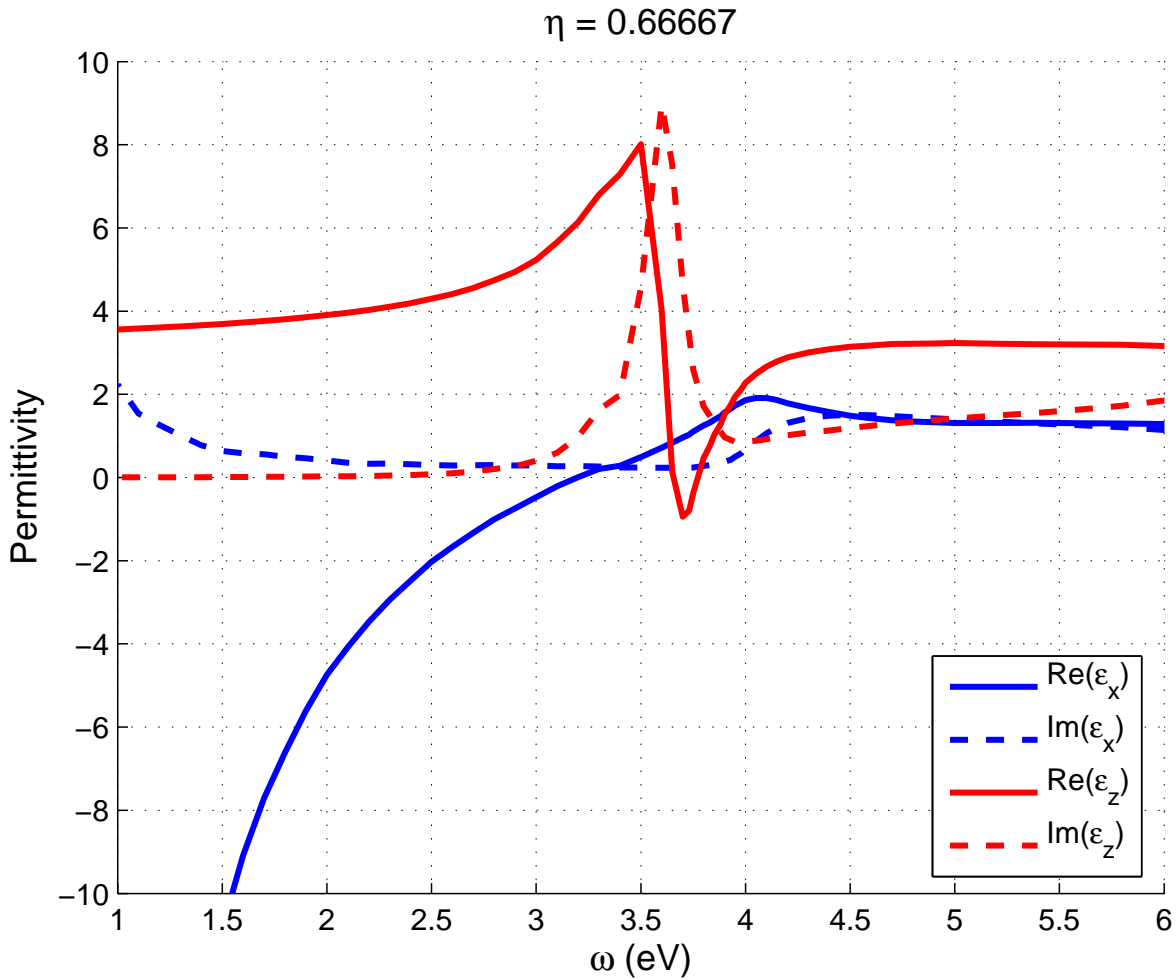,width=\columnwidth}
\caption{(Color online) 
The dielectric permittivity of the metamaterial constructed
from layers of silver and silica. This and the following graph show the
real and imaginary parts of the in-plane and perpendicular components of 
the permittivity for different layer thickness ratios; in this case, $\eta=2/3$,
which means that the layers of silica are one and a half times 
as thick as the layers of metal. 
}
\label{fig:silver-silica-1}
\end{figure}

\begin{figure}
\epsfig{file=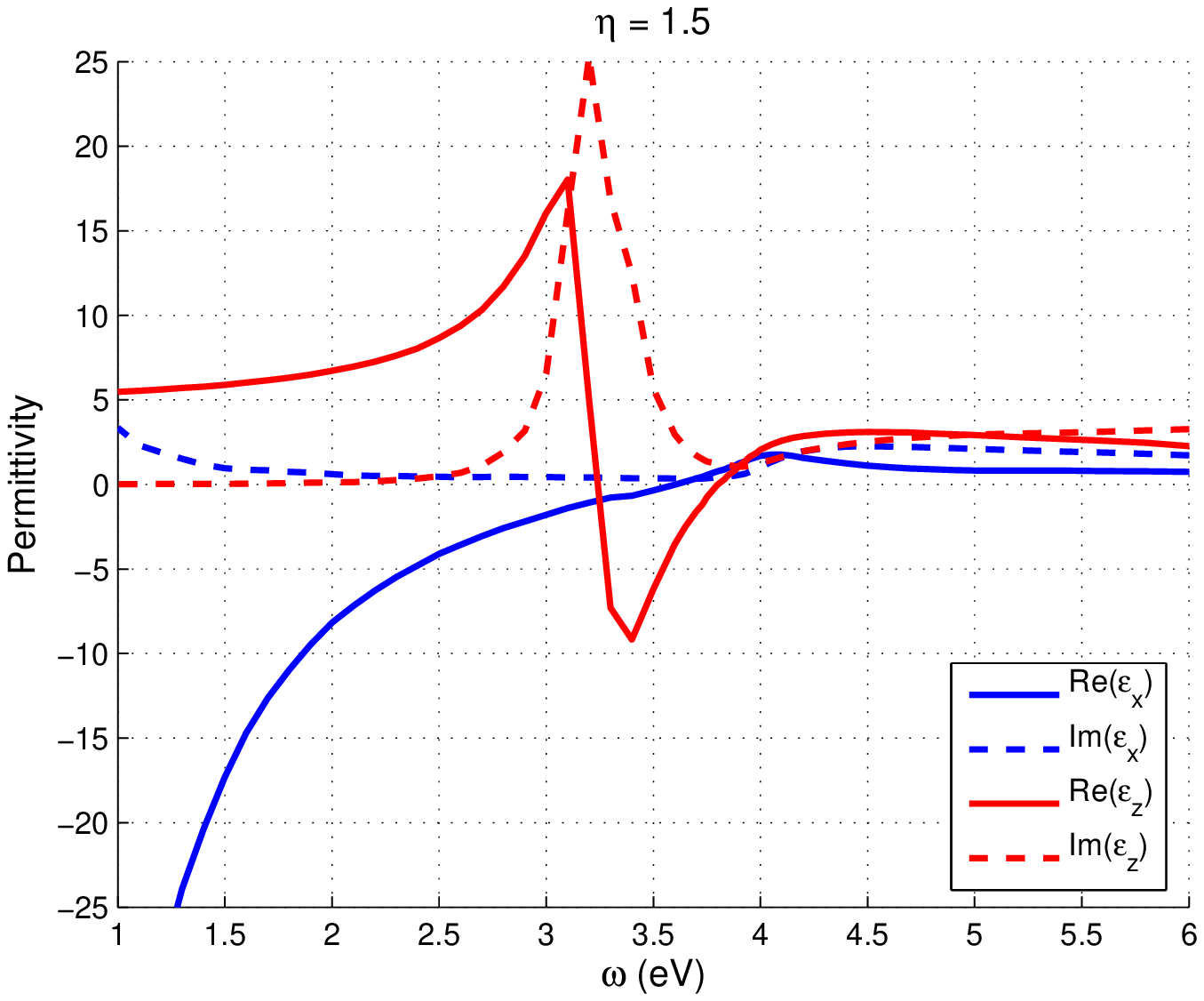,width=\columnwidth}
\caption{(Color online)
The effective permittivity when $\eta=1.5$. The silver layers are now
thicker, and the amount of absorption has increased: the imaginary parts of the
permittivity are now larger. However, the real parts are also correspondingly
larger in magnitude.}
\label{fig:silver-silica-2}
\end{figure}

For a layered metal-dielectric system, we can tune the response either by 
altering the frequency or by changing the ratio of layer thicknesses.
\citep{yu:optical_transmittance_ag_ox} This is
demonstrated by figures \ref{fig:silver-silica-1} 
and \ref{fig:silver-silica-2}, which 
show the real and imaginary parts of the effective permittivity for two 
different thickness ratios, for a system composed of alternating layers of
silver and silica.
The material data from  which these plots are constructed have been
taken from the books by 
\citet{palik:optical_constants} and \citet{nikogosyan:properties}.
In both graphs, there are two regions in which $\Re(\epsilon_x)$ and 
$\Re(\epsilon_z)$ take opposite signs. In the first region, 
which includes energies up to approximately 3.2 eV,
$\Re(\epsilon_x)$ is negative; in the second, which 
consists of a small range of energies around 3.6 eV, $\Re(\epsilon_x)$ is
positive. 

By choosing a suitable value of $\eta$, we can make the real parts
of $\epsilon_z$ and $\epsilon_x$ take opposite signs over a 
range of frequencies. We 
investigate the consequences of this in the next section.

\section{Permittivity with direction-dependent sign}
\label{sec:permittivity}
The unusual behavior of the layered materials can be understood by considering
the dispersion relation between the frequency $\omega$ and the wave vector
${\mathbf k}$. We assume that we are dealing with non-magnetic materials, so 
that the magnetic permeability $\mu=1$. If the dielectric permittivity is 
anisotropic, the interesting waves are those with transverse 
magnetic (TM) polarization. The dispersion relation for these waves is
\begin{equation}
\frac{k_x^2}{\epsilon_z}+\frac{k_z^2}{\epsilon_x}=\frac{\omega^2}{c^2} =
k_0^2.
\label{eq:dispersion_aniso}
\end{equation}
We have taken $k_y$ to be zero, since the $x$- and $y$-directions are 
equivalent.
When $\epsilon_x$ and $\epsilon_z$ are both positive, the relationship 
between $k_x$ and $k_z$ is similar to that in free space: for small $k_x$, 
$k_z$ is real, but when $k_x$ becomes large, $k_z$ becomes imaginary. The
propagation of the wave in the $z$-direction is governed by $k_z$; when
$k_z$ is imaginary, the wave is evanescent: it decays exponentially with $z$.

However, when $\epsilon_x$ and $\epsilon_z$ have opposite signs, $k_z$ is real
for a much wider range of values of $k_x$. 
Even the high spatial frequency components with
large $k_x$, which would normally be evanescent, now correspond to real values
of $k_z$, and hence to propagating waves. 

\begin{figure}
\epsfig{file=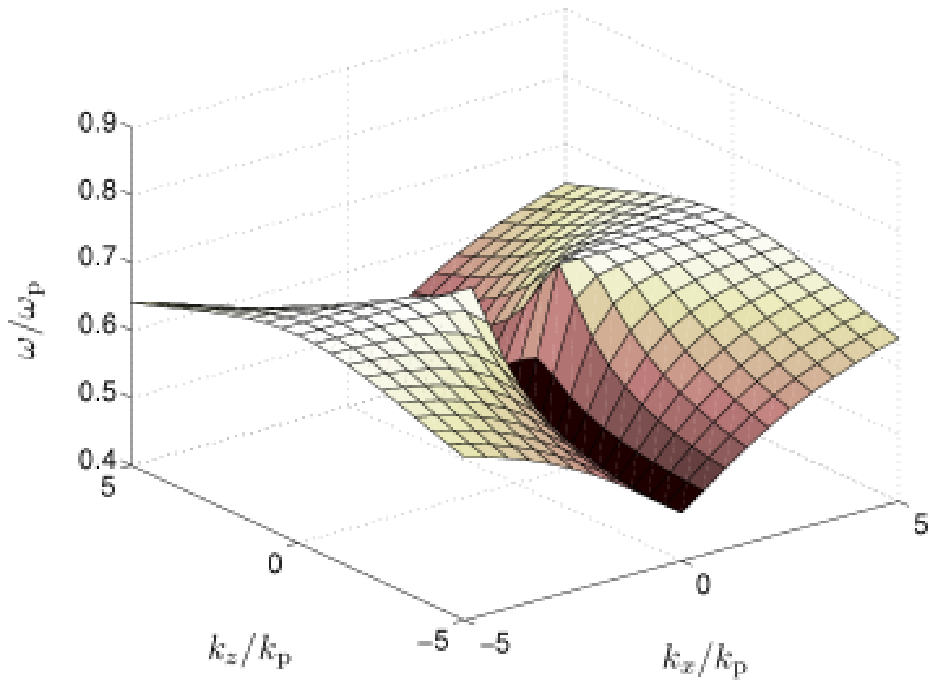,width=0.7\columnwidth}
\epsfig{file=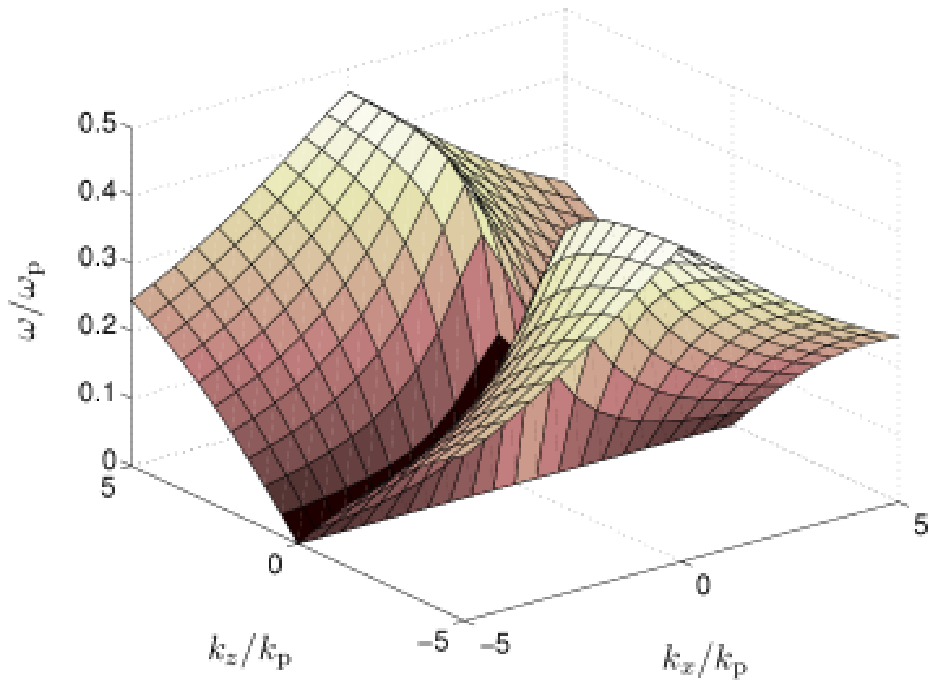,width=0.7\columnwidth}
\caption{(Color online) 
The dispersion relation for an idealized metal-insulator system.
The permittivity of the metal is given by \eqref{eq:eps_metal} with
$\epsilon_\mathrm{m}(\infty)=2.0$, while the dielectric has permittivity
$\epsilon_\mathrm{d}=2.5$; the layers are of equal width ($\eta=1$). 
$k_\mathrm{p}$ is the wave vector corresponding to
the plasma frequency ($k_\mathrm{p}=\omega_\mathrm{p}/c$).
The first two bands are shown; they
have been separated to make visualisation easier, but there is no band gap. 
The plots are symmetric about the planes $k_x=0$ and $k_z=0$. 
}
\label{fig:dispersion}
\end{figure}

If we want to plot the dispersion relation, we have to remember that the
permittivity itself is frequency-dependent. To get an idea of what the 
dispersion relation looks like, we can use an idealized model: we imagine a 
metamaterial whose layers are composed of equal thicknesses of
a dielectric, with positive, 
frequency-independent permittivity, and a metal, with the simple plasma-like
permittivity
\begin{equation}
\epsilon_\mathrm{m}(\omega)=\epsilon_\mathrm{m}(\infty)
 - \frac{\omega_\mathrm{p}^2}{\omega^2}.
\label{eq:eps_metal}
\end{equation}
For now, we assume that the materials are non-absorbing. The resulting
dispersion relation is plotted in figure \ref{fig:dispersion}.
We can identify two distinct bands from the figure. 
In the lower, $\epsilon_x$ is negative, while $\epsilon_z$ is positive;
the signs are reversed in the upper band. In both cases, the contours of
constant $\omega$ are hyperbolae. In the lower band, these hyperbolae 
are centered on 
the $k_x$-axis, while in the upper, they are centered on the $k_z$-axis.

In fact, there is also a third band at
high frequencies, but this is the least interesting regime and is not shown 
in figure \ref{fig:dispersion}: both components of
the permittivity are positive here. 

\begin{figure}
\epsfig{file=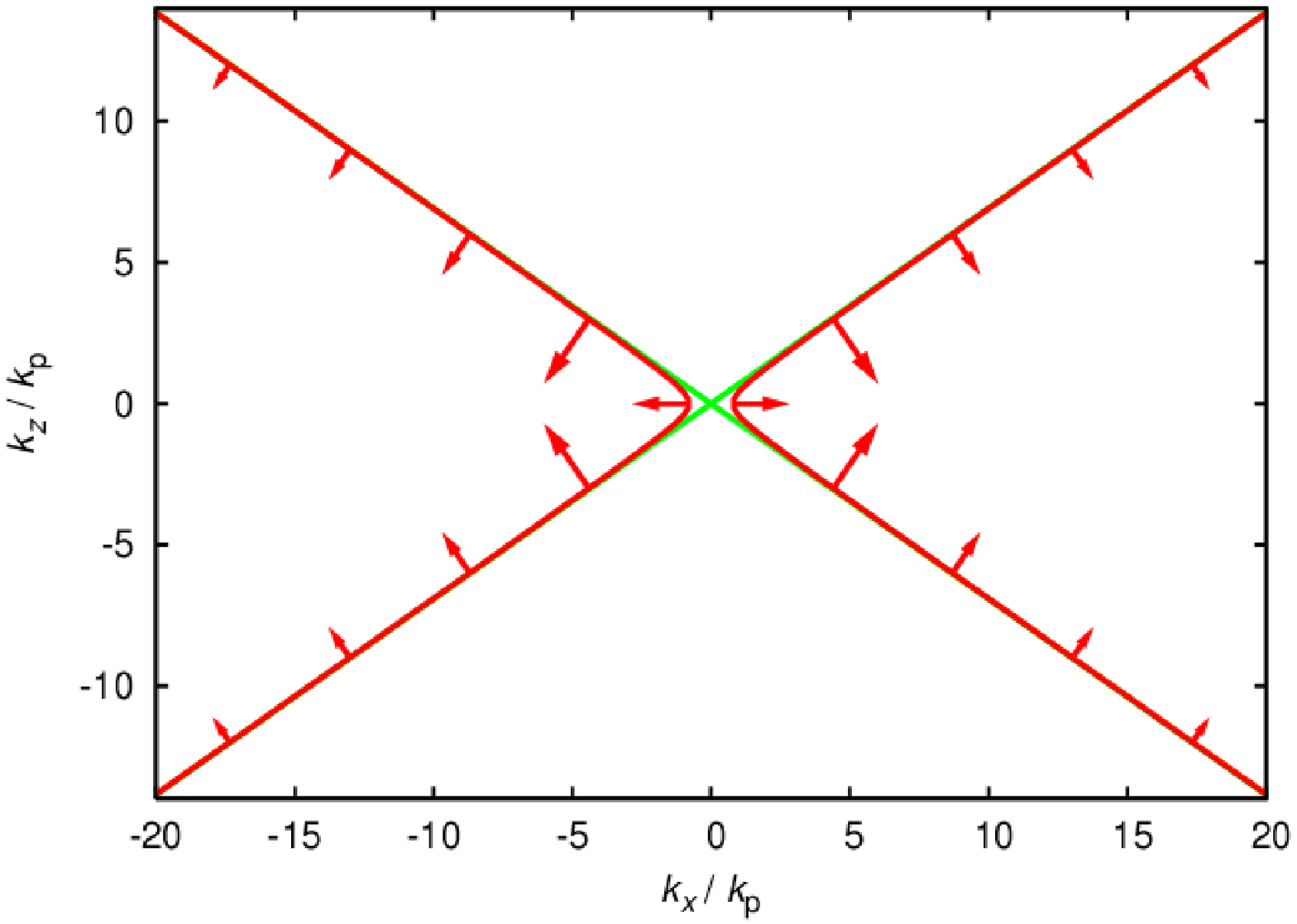, width=\columnwidth}
\caption{(Color online) The relationship between $k_x$ and $k_z$ 
for $\omega=0.2\omega_\mathrm{p}$ (in the 
middle of the lower band in figure \ref{fig:dispersion}). The straight lines
show the asymptotes given by equation \eqref{eq:asymptote}. The group velocity
is indicated by the arrows, which are perpendicular to the curve; the length of
the arrows is proportional to the magnitude of the group velocity.}
\label{fig:group-velocity}
\end{figure}

The dispersion relation also provides the key to the preferred propagation
direction. This is determined by the group velocity. A constant-frequency 
section of the dispersion relation (cut across the first band) is plotted in
figure \ref{fig:group-velocity}. 
The hyperbolic form of the curve means that
for large $|k_x|$, it tends to the following straight line:
\begin{equation}
k_z=\sqrt{-\frac{\epsilon_x}{\epsilon_z}} |k_x|.
\label{eq:asymptote}
\end{equation}
The group velocity is perpendicular to the constant-$\omega$ contours like the
one plotted in figure \ref{fig:group-velocity}. The figure demonstrates that
apart from a region around $k_z=0$, the group velocity vectors all point in
almost the same two directions: this is the basis for the preferred direction
of propagation. Remembering that the $x$- and $y$-directions are equivalent,
we can see that the preferred directions form a cone around the $z$-axis. The
half-angle of the cone is 
\begin{equation}
\theta=\arctan\sqrt{-\frac{\epsilon_x}{\epsilon_z}}.
\label{eq:angle}
\end{equation}
In the region around $k_z=0$, the
arrows point outside the cone. 
In this band, there are no propagating 
modes in a small region around $k_x=0$, and no propagating modes with a group
velocity vector lying inside the cone. 

If we take a cross-section from the second band, instead of the first, 
we also see a hyperbolic contour; the plot resembles figure 
\ref{fig:group-velocity}, but rotated by 90$^\circ$. The group velocities for 
the modes around $k_x=0$ now point inside the cone, rather than outside.

\begin{figure}
\begin{center}
\epsfig{file=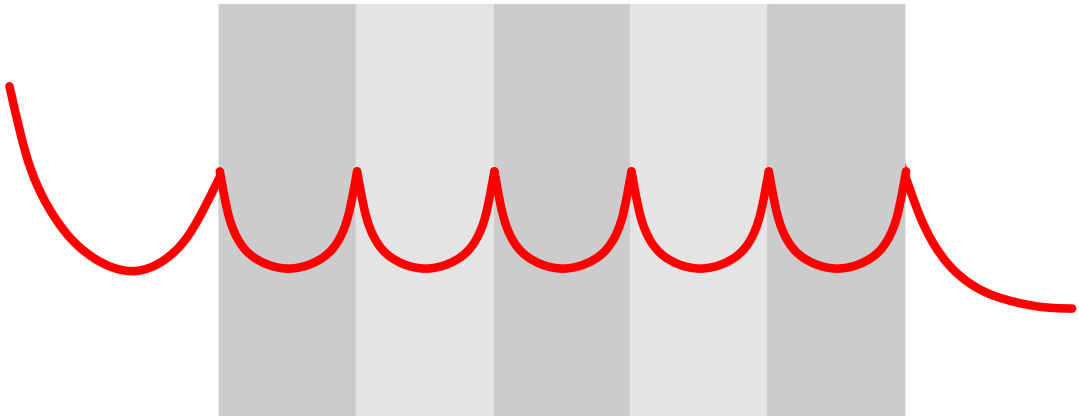,width=0.8\columnwidth}
\end{center}
\caption{(Color online) Schematic diagram of the  
transmission of normally evanescent waves, showing the role of
surface plasmons. The line represents the electric field strength.}
\label{fig:coupled-plasmons}
\end{figure}

To conclude this section, we look at the physical process that allows our
layered metamaterial to mimic an anisotropic material and to support the 
propagation of normally evanescent waves. The key fact is that surface plasmons
are supported at an interface where the permittivity changes sign. When the
metal permittivity is negative, this sign change occurs at every interface;
the wave is transmitted via coupled surface plasmons, as indicated in figure
\ref{fig:coupled-plasmons}. 

\section{Transmission through an anisotropic system}
\label{sec:transmission}
We have seen that we can produce a metamaterial with interesting properties by
stacking alternating layers of metal and dielectric. Next, we look at a
slab of this material, and examine the transmission coefficient.

We assume that the slab is embedded in a uniform medium of constant 
permittivity (which may be unity, representing vacuum). 
In such a medium, the dispersion relation \eqref{eq:dispersion_aniso} becomes
\begin{equation}
{k_x^2}+{k_z'^2}=k_0^2\epsilon .
\label{eq:dispersion_iso}
\end{equation}
We write $k_z'$ to distinguish the $z$-component of the wave vector in the 
surrounding medium from that in the slab. 
The transmission coefficient for TM waves is
\begin{equation}
t(k_x, \omega)= 
\frac{2}{2\cos k_z d - i\left( \frac{k_z'\epsilon_x}{k_z \epsilon}
+\frac{k_z \epsilon}{k_z'\epsilon_x}
\right) \sin k_z d}
\label{eq:transmission}
\end{equation}
where the dispersion relations \eqref{eq:dispersion_aniso}
and \eqref{eq:dispersion_iso}
are used to define $k_z$ and $k_z'$ in terms of $k_x$ and $\omega$. 

\begin{figure}
\epsfig{file=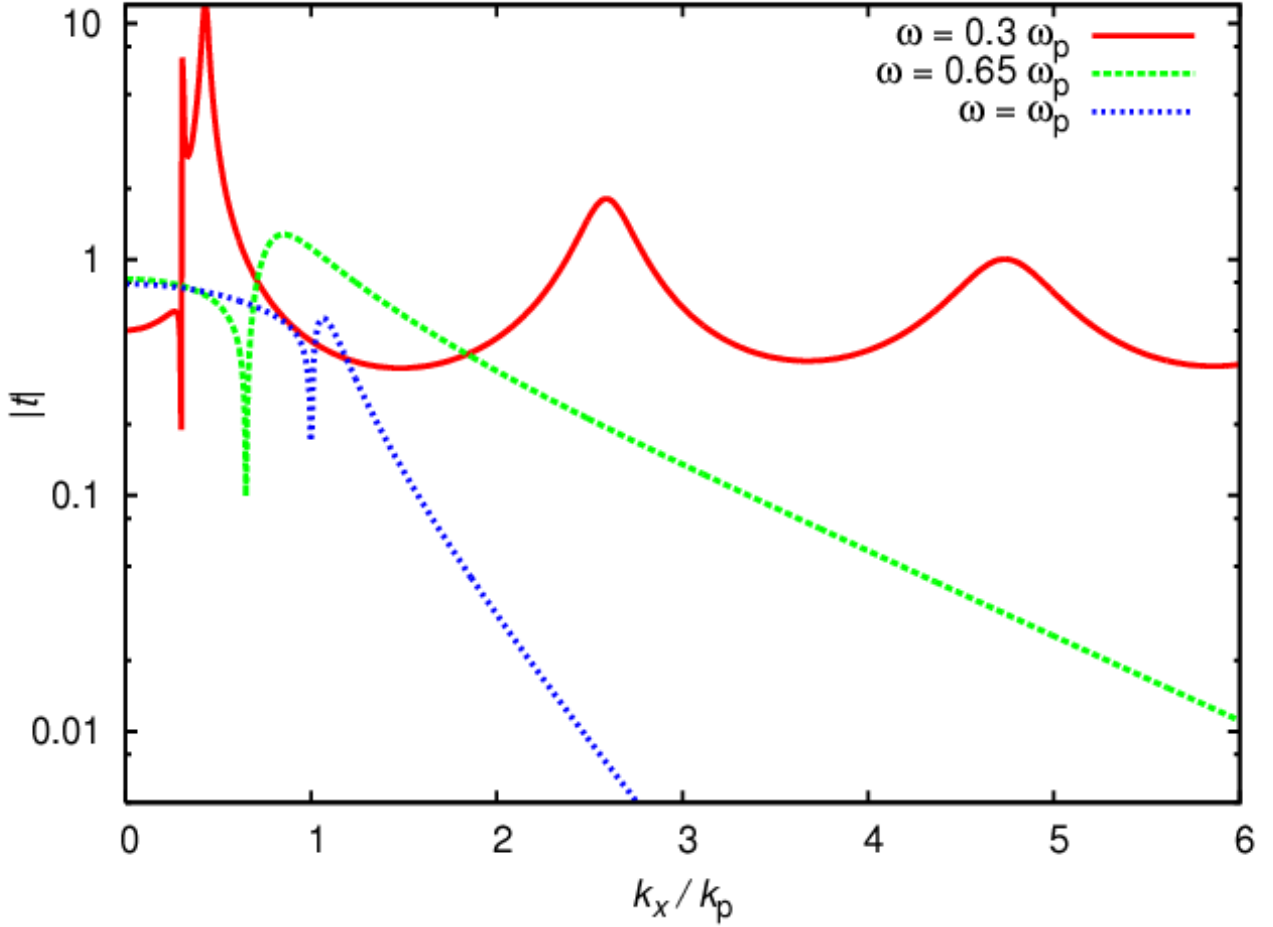, width=\columnwidth}
\caption{(Color online) The transmission coefficient, as defined in equation 
\eqref{eq:transmission}, with the surrounding medium taken to be air. 
The slab width used is $d=2/k_\mathrm{p}$, which would be of the 
order of 100nm for a plasma frequency $\omega_\mathrm{p} \sim 4$eV. The results for
three different frequencies are plotted, corresponding to the three bands 
referred to in the discussion of figure \ref{fig:dispersion}. The real part of
$\epsilon_x$ is negative when $\omega=0.3\omega_\mathrm{p}$, while that of 
$\epsilon_z$ is 
positive; the signs are reversed when $\omega=0.65\omega_\mathrm{p}$. At higher 
frequencies, both are positive. The permittivity of the metal is taken to be
$\epsilon_\mathrm{m}=1.7+0.6i-\omega_\mathrm{p}^2/\omega^2$, while
that of the dielectric is $\epsilon_\mathrm{d}=2.5$.}
\label{fig:transmission}
\end{figure}

In figure \ref{fig:transmission}, we plot the transmission coefficient for 
three different regimes, corresponding to the three different frequency ranges.
At high frequencies (here represented by $\omega=\omega_\mathrm{p}$), 
both components of the metamaterial permittivity are 
positive. In this regime, the transmission coefficient is close to unity for
small wave vectors. It drops abruptly to zero
at $k_x=k_0$, and rises equally sharply afterward, again approaching 
unity; finally, it decays exponentially for larger
wave vectors. Very similar behavior is observed in the
intermediate frequency range ($\omega=0.65\omega_\mathrm{p}$). This time, the maximum
following the zero at  $k_x=k_0$ is higher, 
and the rate of exponential decay for large
$k_x$ is less rapid.

The most interesting frequency range is the lowest one ($\omega=0.3\omega_\mathrm{p}$). 
There is the usual zero in the transmission at $k_x=k_0$, followed by a 
very sharp peak. However,
there is also significant transmission even for large wave vectors; the
transmission coefficient has a series of peaks, decreasing in magnitude, and
approximately periodic in $k_x$. 
The resonances correspond to localized states for the
slab; they are in turn antisymmetric and symmetric. There is a difference 
between the first two resonances (just above $k_x=k_0$) and those for
higher wave vectors. For the first two, the wave is non-propagating inside the
slab (because $k_z$ is almost purely imaginary): the resonances therefore
consist of coupled surface plasmons located on each
surface of the slab. For the higher wave vectors, the wave is able to propagate 
\footnote{The idea of a propagating wave inside the 
metamaterial applies in the effective medium approximation; in the microscopic 
picture, the ``propagating'' wave is made up of coupled evanescent waves, as 
indicated in figure \ref{fig:coupled-plasmons}.} 
within the slab (because $k_z$ is almost purely real), and the transmission 
peaks correspond to Fabry-Perot resonances --
standing waves inside the slab. \citep{yu:optical_transmittance_ag_ox}

In fact, a similar set of peaks would be visible in the intermediate-frequency
regime, were it not for absorption. The material parameters used to 
generate figure \ref{fig:transmission} include a realistic amount of absorption,
and a glance at figures \ref{fig:silver-silica-1} and \ref{fig:silver-silica-2}
shows that absorption is high in the region where $\Re(\epsilon_z)$ becomes 
negative. The localized states are supported in both the low- and 
intermediate-frequency ranges, but are suppressed in the latter by high
absorption. 

\section{Imaging a line source}

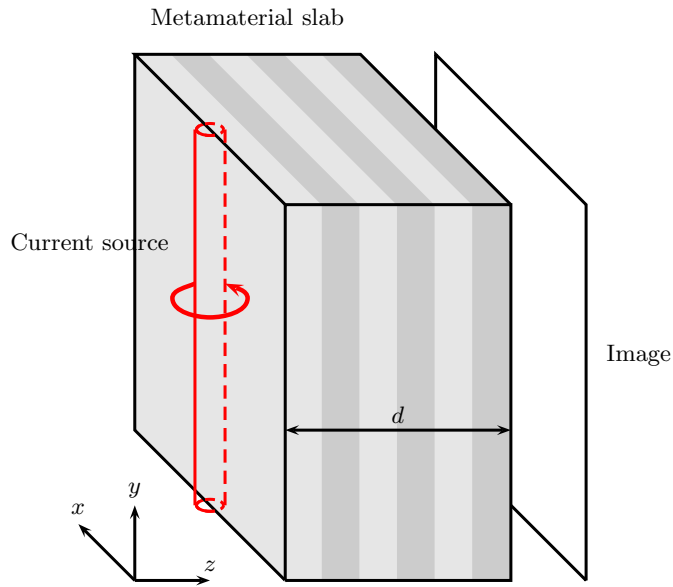
\begin{figure}
\begin{center}
\begin{pspicture}(2,0)(8,8)
\newrgbcolor{lightgrey}{0.8 0.8 0.8}
\newrgbcolor{vlightgrey}{0.9 0.9 0.9}
\psline[linecolor=black,linewidth=0.4mm](8.0,0.0)(6.0,2.0)(6.0,7.0)(8.0,5.0)(8.0,0.0)
\psellipse[linecolor=red,fillstyle=none,linewidth=0.4mm](3.0,1.0)(0.2,0.1)
\pspolygon[linecolor=vlightgrey,fillstyle=solid,fillcolor=vlightgrey](4,0)(2,2)(2,7)(4,5)
\psline[linecolor=black,linewidth=0.4mm]{-}(4,0)(2,2)(2,7)
\psellipse[linecolor=red,fillstyle=none,linewidth=0.4mm](3.0,6.0)(0.2,0.1)
\psline[linecolor=red,linewidth=0.4mm]{-}(2.8,6)(2.8,1)
\pspolygon[linecolor=vlightgrey,fillstyle=solid,fillcolor=vlightgrey](2.0,7)(2.5,7)(4.5,5)(4.5,0)(4.0,0)(4.0,5)
\pspolygon[linecolor=lightgrey,fillstyle=solid,fillcolor=lightgrey](4.5,0)(5,0)(5,5)(3,7)(2.5,7)(4.5,5)
\pspolygon[linecolor=vlightgrey,fillstyle=solid,fillcolor=vlightgrey](5.0,0)(5.5,0)(5.5,5)(3.5,7)(3.0,7)(5,5)
\pspolygon[linecolor=lightgrey,fillstyle=solid,fillcolor=lightgrey](5.5,0)(6,0)(6,5)(4,7)(3.5,7)(5.5,5)
\pspolygon[linecolor=vlightgrey,fillstyle=solid,fillcolor=vlightgrey](6.0,0)(6.5,0)(6.5,5)(4.5,7)(4.0,7)(6.0,5)
\pspolygon[linecolor=lightgrey,fillstyle=solid,fillcolor=lightgrey](6.5,0)(7,0)(7,5)(5,7)(4.5,7)(6.5,5)
\psline[linecolor=black,linewidth=0.4mm]{-}(4,0)(7,0)(7,5)(4,5)(4,0)
\psline[linecolor=black,linewidth=0.4mm]{-}(4,5)(2,7)(5,7)(7,5)
\psellipse[linecolor=red,linestyle=dashed,fillstyle=none,linewidth=0.4mm](3.0,1.0)(0.2,0.1)
\psellipse[linecolor=red,linestyle=dashed,fillstyle=none,linewidth=0.4mm](3.0,6.0)(0.2,0.1)
\psline[linecolor=red,linestyle=dashed,linewidth=0.4mm]{-}(3.2,6)(3.2,1)
\psecurve[linecolor=red,linewidth=0.6mm]{->}(3.0,4.0)(2.8,3.95)(2.5,3.75)(3.0,3.5)(3.5,3.75)(3.2,3.95)(3.0,4.0)
\rput(3.5,7.5){\psframebox[linestyle=none,fillstyle=none]{Metamaterial slab}}
\rput(8.7,3.0){\psframebox[linestyle=none,fillstyle=none]{Image}}
\rput(1.4,4.5){\psframebox[linestyle=none,fillstyle=none]{Current source}}
\psline[linecolor=black,linewidth=0.4mm]{<->}(4,2)(7,2)
\rput(5.5,2.2){\psframebox[linestyle=none,fillstyle=none]{$d$}}
\psline[linecolor=black,linewidth=0.4mm]{->}(2,0)(3,0)
\psline[linecolor=black,linewidth=0.4mm]{->}(2,0)(2,1)
\psline[linecolor=black,linewidth=0.4mm]{->}(2,0)(1.25,0.75)
\rput(3,0.2){\psframebox[linestyle=none,fillstyle=none]{$z$}}
\rput(2.0,1.2){\psframebox[linestyle=none,fillstyle=none]{$y$}}
\rput(1.25,0.95){\psframebox[linestyle=none,fillstyle=none]{$x$}}
\end{pspicture}
\end{center}
\caption{(Color online) Imaging a solenoidal line source.}
\label{fig:line_source}
\end{figure}

We have seen that the layered system allows enhanced transmission of 
high-spatial-frequency components at certain frequencies. This gives us hope
that we may achieve sub-wavelength imaging using the slab 
As a test, we 
consider the image of the line source pictured in figure \ref{fig:line_source}.

In the absence of the metamaterial, the field generated by this source is
\begin{equation}
{\mathbf E}({\mathbf r}) = \int_{-\infty}^\infty \int_{-\infty}^\infty 
e^{ik_xx+ik_yy+ik_z'z-i\omega t} \tilde{E}_0(k_y) \left(\hat{\mathbf x}-
\frac{k_x}{k_z'}\hat{\mathbf z}\right)\;dk_x\;dk_y,
\end{equation}
where the current profile in the $y$-direction is as yet unspecified.
As before, $k_z'$ represents the $z$-component of the wave vector in the
surrounding medium.

\begin{figure}
\epsfig{file=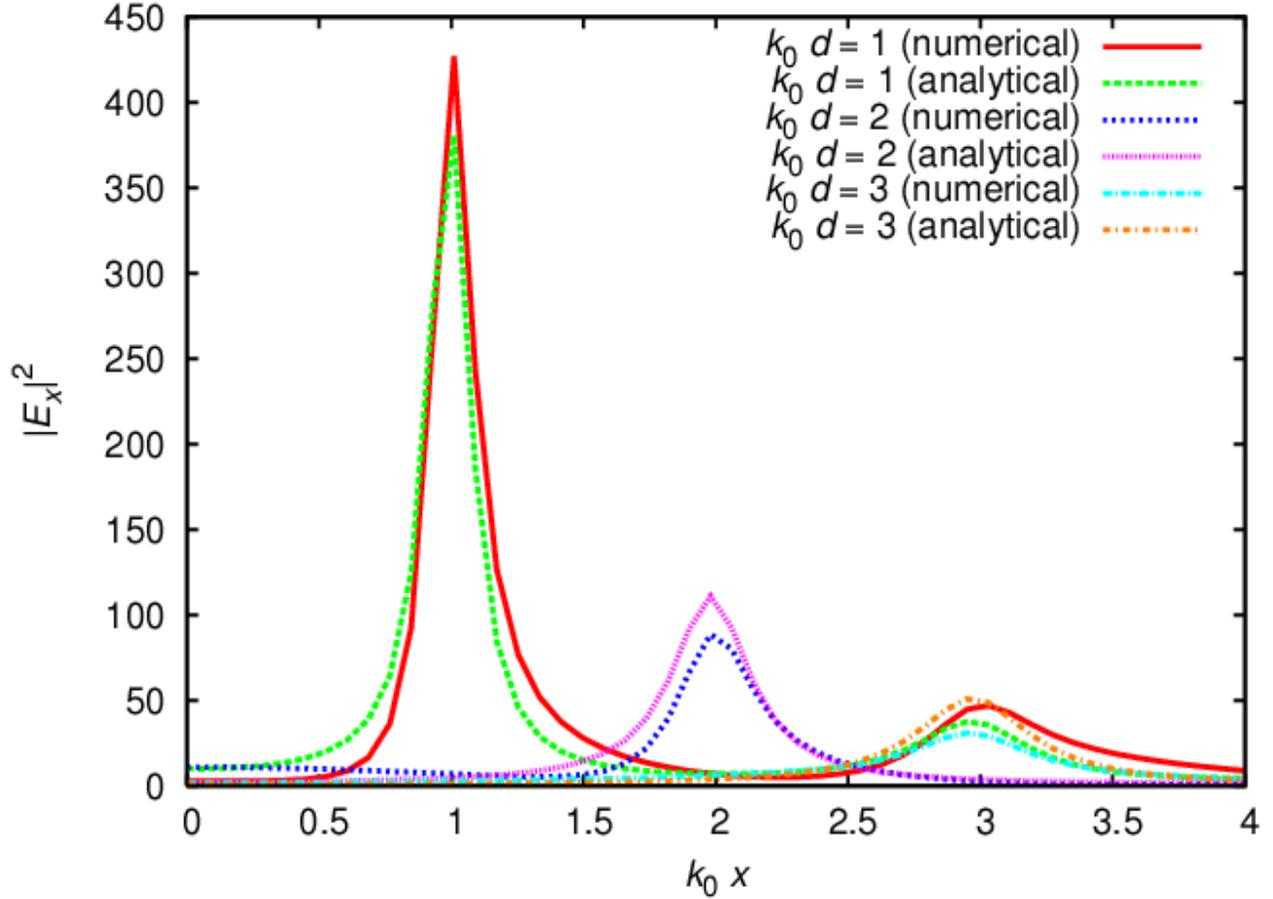, width=\columnwidth}
\caption{(Color online) 
The transmitted field intensity, comparing analytical and numerical 
results. The parameters used were $\epsilon_\mathrm{d}=2.5$, $\epsilon_\mathrm{
m}=1.7+0.05i-\omega_\mathrm{p}/\omega^2$, with $\omega=0.68\omega_\mathrm{p}$. In each case, we
plot the field just beyond the slab (so that for example when 
$k_0d=2.0$, $k_0z=2.01$).}
\label{fig:transmitted_field_sheet_2}
\end{figure}

When we place the metamaterial next to the source, as shown in the figure,
some radiation will be reflected from the slab and will generate additional
currents. If we neglect these, we can estimate the $x$-component of 
the transmitted field as
\begin{equation}
E_x^\mathrm{TM}({\mathbf r}) = \int_{-\infty}^\infty \int_{-\infty}^\infty
e^{ik_x x+ik_y y+ik_z'(z-d)-i\omega t} 
t\left(\sqrt{k_x^2+k_y^2},\omega\right)
\cdot
\frac{\tilde{E}_0(k_y) k_x^2}{k_x^2+k_y^2} \;dk_x\;dk_y.
\end{equation}
Note that this is the TM component of the field.  In general, there will also 
be a TE component which must be calculated. However, if we 
consider a line source which is uniform in strength and infinitely long, so 
that $E_0(k_y) \propto \delta(k_y)$, the entire field is transverse magnetic. 
In this case, the calculation reduces to the solution of the following integral:
\begin{equation}
E_x^\mathrm{TM}({\mathbf r}) = \frac{E_0}{k_0} \int_{-\infty}^\infty 
\frac{2e^{ik_x x+ik_z'(z-d)-i\omega t}}{2\cos k_zd -i\left(\frac{k_z'
\epsilon_x}{k_z}+\frac{k_z}{k_z'\epsilon_x}\right)\sin k_zd } \;dk_x.
\end{equation}
The integral can be solved approximately when the frequency is in the 
intermediate range: that is, when $\Re(\epsilon_x)>0$ and $\Re(\epsilon_z)<0$.
The resonant states then all have large $k_x$; they are the standing wave
states discussed in the previous section, rather than the coupled surface 
plasmon states (which have $k_x$ close to $k_0$). We are therefore justified in 
making the near field approximation, which leads to the following analytic form 
for the $x$-component of the transmitted field:
\begin{equation}
E_x^{\mathrm{TM}}({\mathbf r}) \approx \frac{\pm 4\pi i E_0 }{k_0  d\left(1/
\epsilon_z-\epsilon_x\right)} \cdot \frac{e^{-ik_1\left(|x|-i(z-d)\right)-i
\omega t}}{1+e^{-i\Delta k \left(|x|-i(z-d)\right)}} ,
\end{equation}
where
\begin{equation}
k_1 =\frac{1}{d}\sqrt{-\frac{\epsilon_z}{\epsilon_x}}\arctan 
\left(\frac{2\sqrt{-\epsilon_x\epsilon_z}}{1+\epsilon_x\epsilon_z}\right)
\end{equation}
and 
\begin{equation}
\Delta k = \frac{\pi}{d}\sqrt{-\frac{\epsilon_z}{\epsilon_x}}.
\end{equation}
In this approximation, $E_x^\mathrm{TM}$ and $E_z^\mathrm{TM}$ are
identical to within a phase factor. In figure 
\ref{fig:transmitted_field_sheet_2}, we plot the intensity of the transmitted
field, comparing the approximate analytical solution with the results of 
numerical integration. 
To generate the plot, we take an unrealistically low value for the absorption in
the metal; the point of the graph is to compare numerical and analytical 
results, but also to demonstrate the features which we hope to be able to 
observe. 

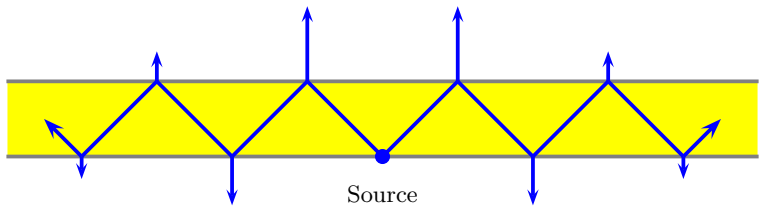
\begin{figure}
\begin{center}
\begin{pspicture}(0,0)(12,3)
\psframe[linestyle=none,fillstyle=solid,fillcolor=yellow](1,1)(11,2)
\psline[linecolor=gray,linewidth=0.5mm]{-}(1,1.0)(11,1.0)
\psline[linecolor=gray,linewidth=0.5mm]{-}(1,2.0)(11,2.0)
\psline[linecolor=blue,linewidth=0.5mm]{-}(6,1.0)(7,2.0)
\psline[linecolor=blue,linewidth=0.5mm]{-}(6,1.0)(5,2.0)
\psline[linecolor=blue,linewidth=0.5mm]{-}(8,1.0)(7,2.0)
\psline[linecolor=blue,linewidth=0.5mm]{-}(4,1.0)(5,2.0)
\psline[linecolor=blue,linewidth=0.5mm]{-}(8,1.0)(9,2.0)
\psline[linecolor=blue,linewidth=0.5mm]{-}(4,1.0)(3,2.0)
\psline[linecolor=blue,linewidth=0.5mm]{-}(10,1.0)(9,2.0)
\psline[linecolor=blue,linewidth=0.5mm]{-}(2,1.0)(3,2.0)
\psline[linecolor=blue,arrowsize=2mm,linewidth=0.5mm]{->}(10,1.0)(10.5,1.5)
\psline[linecolor=blue,arrowsize=2mm,linewidth=0.5mm]{->}(2,1.0)(1.5,1.5)
\pscircle[fillstyle=solid,fillcolor=blue,linecolor=blue](6,1){0.1}
\rput(6.0,0.5){\psframebox[linestyle=none,fillstyle=none]
{Source}}
\psline[linecolor=blue,linewidth=0.5mm]{->}(7,2)(7,3)
\psline[linecolor=blue,linewidth=0.5mm]{->}(5,2)(5,3)
\psline[linecolor=blue,linewidth=0.5mm]{->}(9,2)(9,2.4)
\psline[linecolor=blue,linewidth=0.5mm]{->}(3,2)(3,2.4)
\psline[linecolor=blue,linewidth=0.5mm]{->}(8,1)(8,0.35)
\psline[linecolor=blue,linewidth=0.5mm]{->}(4,1)(4,0.35)
\psline[linecolor=blue,linewidth=0.5mm]{->}(10,1)(10,0.7)
\psline[linecolor=blue,linewidth=0.5mm]{->}(2,1)(2,0.7)
\end{pspicture}
\end{center}
\caption{(Color online)
A schematic showing that reflections lead to periodically-repeated 
images of the two principal peaks.}
\label{fig:reflections}
\end{figure}

First, we note that the position of the peaks is proportional to the slab width.
This is a manifestation of the preferred direction of propagation; within the
metamaterial, the light travels at a fixed angle to the $z$-axis, in the
$xz$-plane (since we have translational invariance in the $y$-direction). 
The secondary peaks which are visible when $k_0d=1.0$ are caused by reflection
from the boundaries; this is why they overlap precisely with the primary 
peaks for the slab with $k_0d=3.0$. The reflections are illustrated in figure
\ref{fig:reflections}.

\begin{figure}
\begin{center}
\epsfig{file=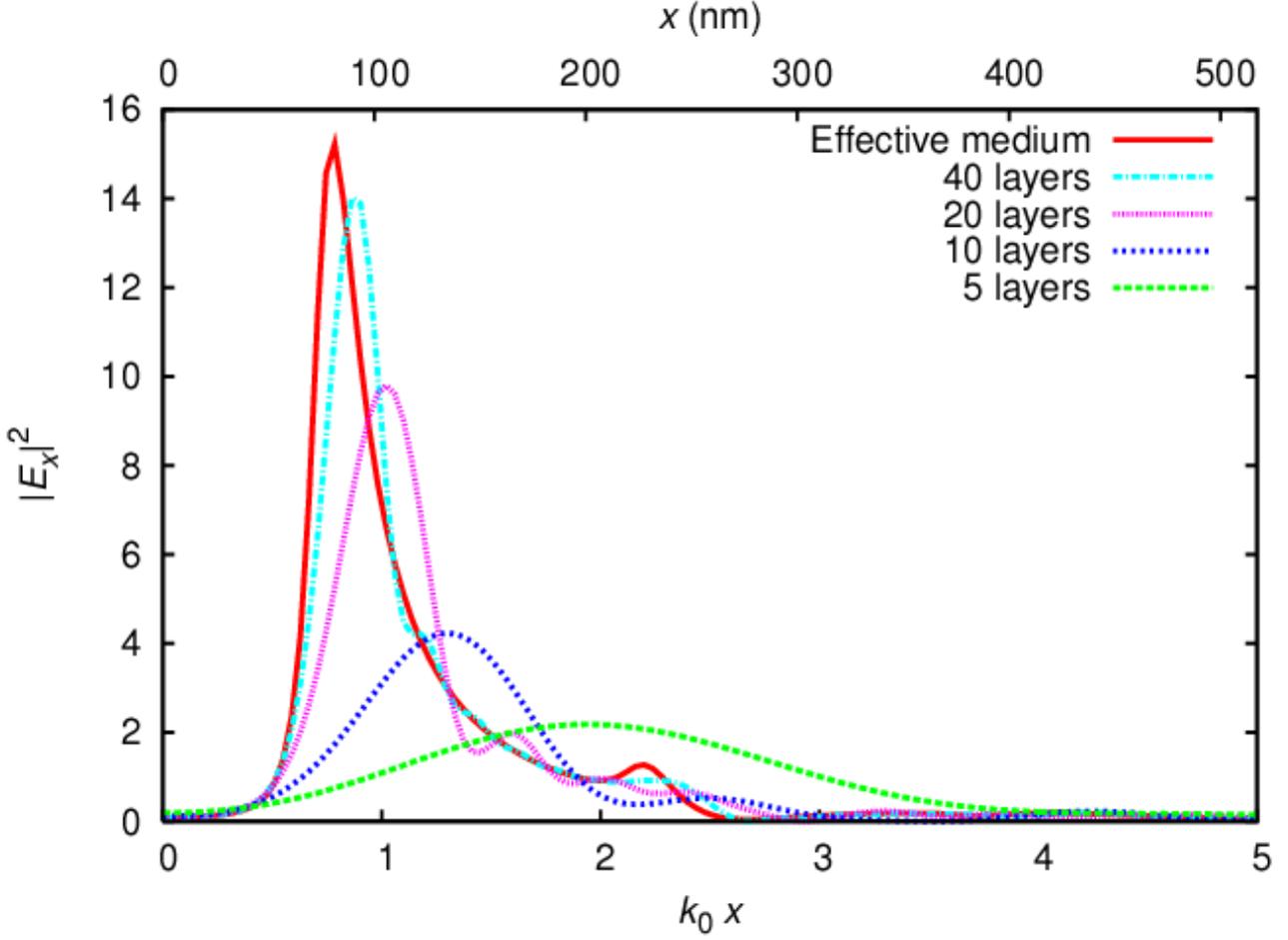, width=\columnwidth}
\caption{(Color online)
The transmitted electric field intensity for a line source, imaged by
a metamaterial slab of thickness $1/k_0$. The material parameters used 
correspond to layers of Ag and ZnS-SiO$_2$, embedded in crystalline 
Ge$_2$Sb$_2$Te$_5$ (a phase-change material used in optical storage devices), 
for light of wavelength 650nm. This corresponds to a total slab width of around
105nm.}
\label{fig:subwavelength}
\end{center}
\end{figure}

In the first frequency regime, the approximate analytical solution is more 
difficult to obtain: there are the additional surface plasmon resonances
close to $k_x=k_0$, for which one cannot make the near field approximation. 
However, it is still possible to obtain numerical results.
As one would expect from figure \ref{fig:transmission}, these are much more
promising: using realistic parameters, we
are able to produce a sharp image, as shown by the line marked 
``Effective medium'' in figure \ref{fig:subwavelength}.
The width of the principal peak in the effective medium approximation
is around $\lambda/10$. Figure \ref{fig:subwavelength} also illustrates the 
results of a more detailed analysis, which goes beyond the simplified 
effective medium approach; we will discuss these next.

\section{Beyond the effective medium approximation}
Treating the layered system as an effective medium is a helpful simplification, 
in terms of both understanding and performing simulations. However, it has
limitations. In this section, we model the system in more detail, considering
the finite width of the layers; naturally, as 
the layers are made thinner, we see that the effective medium 
approximation becomes more appropriate.

\begin{figure}
\begin{center}
\epsfig{file=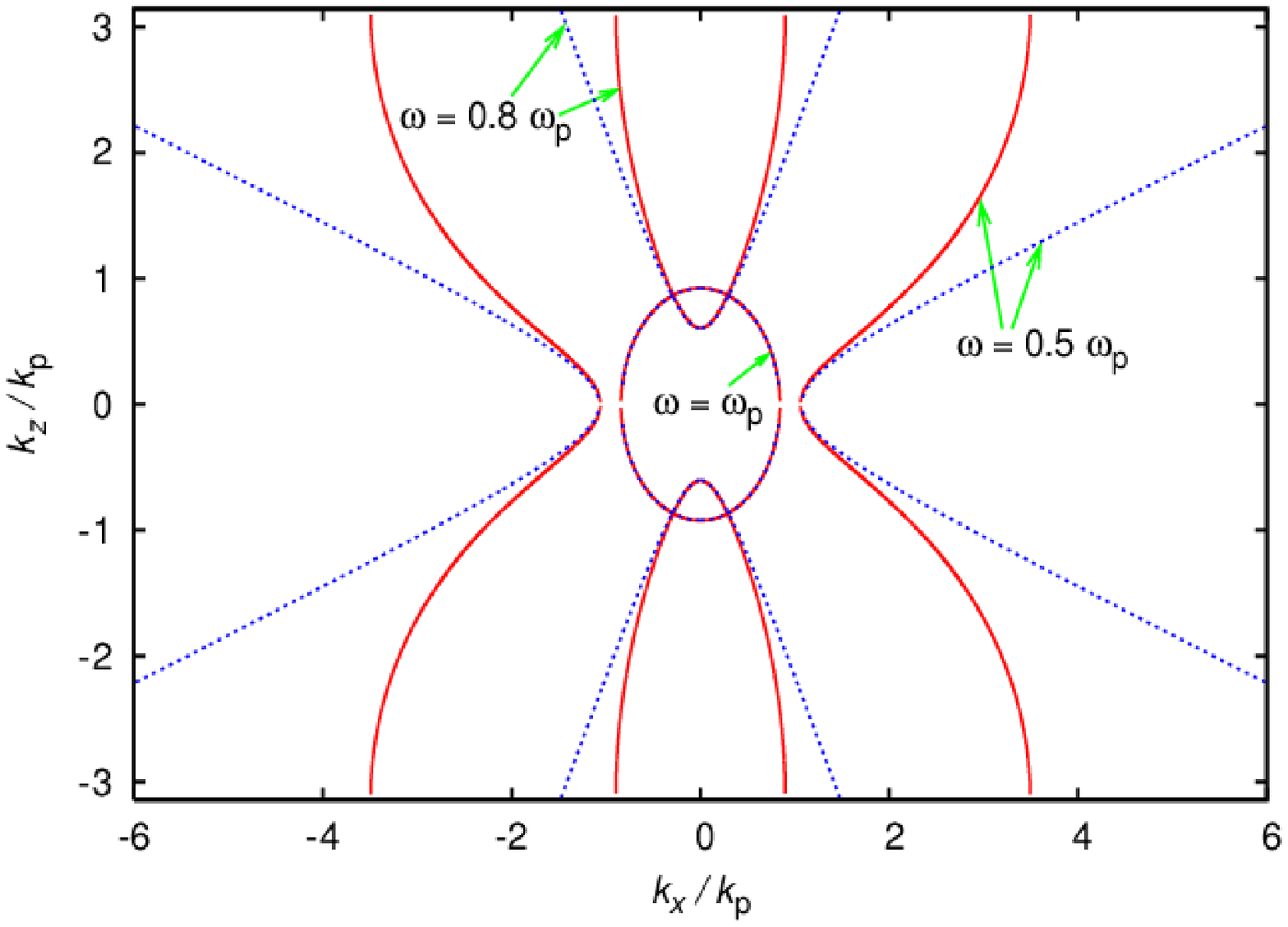, width=\columnwidth}
\caption{(Color online) Isofrequency contours demonstrating the effect of finite
layer width. The solid lines are the contours for a system where the cell
size ($d_1+d_2$) is $1/k_\mathrm{p}$; the result is a Brillouin zone 
of extent $2\pi$ on the $k_z$-axis. The dashed lines are 
the equivalent contours in the effective medium approximation. The
three sets of contours correspond to the three bands of the dispersion 
relation discussed in section \ref{sec:permittivity}; the first two
are hyperbolic in the effective medium approximation, while the third is 
elliptical. The material parameters are the same as those used to generate 
figure \ref{fig:dispersion}.}
\label{fig:isofreq}
\end{center}
\end{figure}

First, we look at the dispersion relation for the layered 
metamaterial. We can obtain $k_z$ as a function of $k_x$ (at a given frequency)
from the characteristic matrix, as described in appendix \ref{app:homogenize}. 
These isofrequency contours are plotted in figure \ref{fig:isofreq}.
The effective medium has full translational symmetry, but this is broken when
considering the structure of finite-width layers; the system becomes periodic,
and the figure  shows part of the first Brillouin zone, which
extends from $k_z=-\pi k_\mathrm{p}$ to $k_z=\pi k_\mathrm{p}$.
The effective-medium contours are deformed by the
new periodicity, and bend towards the zone boundaries. This introduces a new 
cutoff: for a given frequency, there is a value of $k_x$ above which no
propagating solution exists. This affects the resolution of the lens-like
system.

\begin{figure}
\begin{center}
\epsfig{file=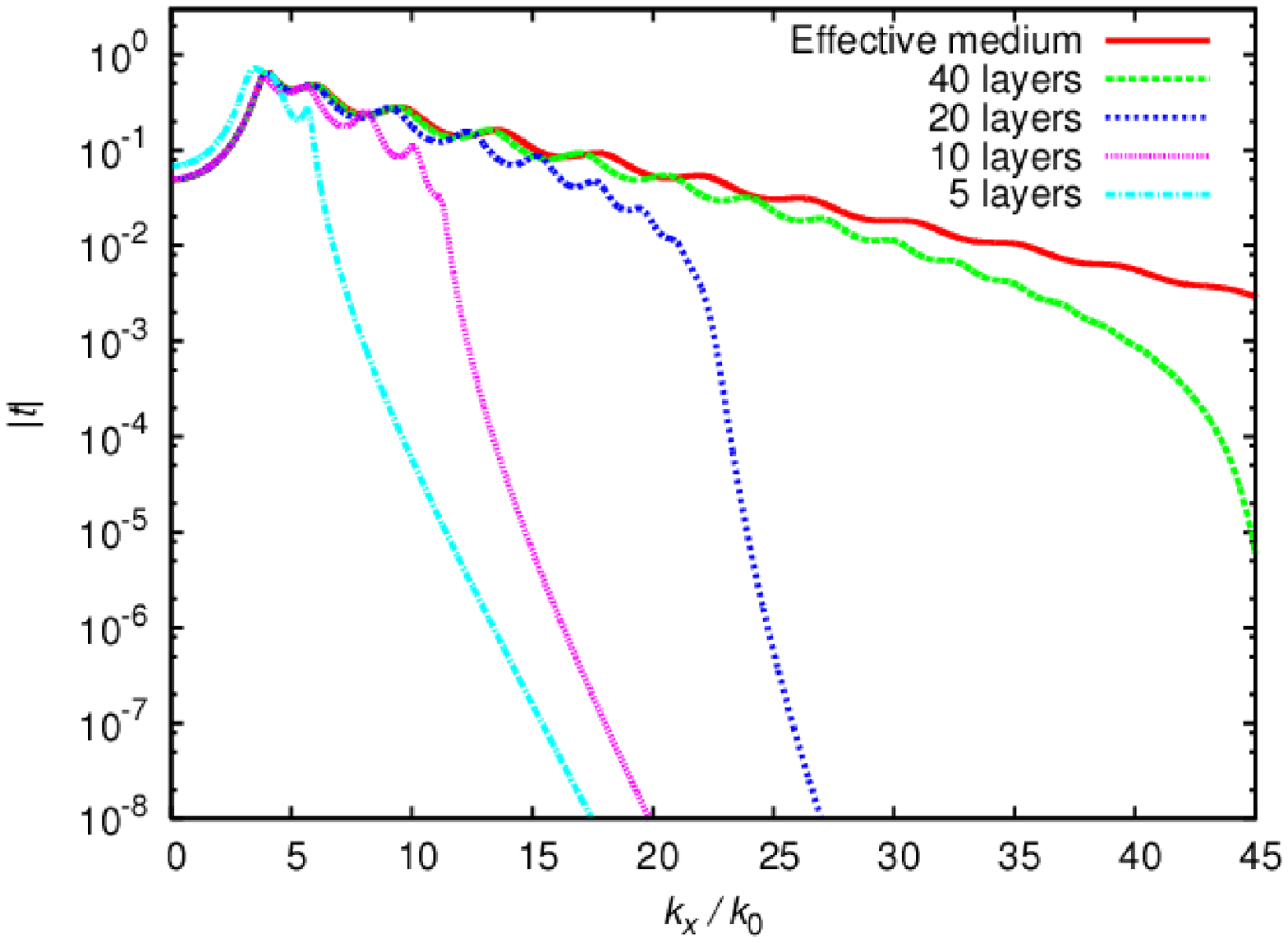, width=\columnwidth}
\caption{(Color online)
The transmission as a function of $k_x$ for various layer widths. The
total slab width is maintained at $1/k_0$ in each case, while the number of 
individual layers is adjusted.}
\label{fig:transmission_layers}
\end{center}
\end{figure}

The next logical step is to investigate the change in the behavior of the 
slab of metamaterial described in section \ref{sec:transmission}. From now on,
we focus on the first band. Figure \ref{fig:transmission_layers} shows that 
the new cutoff in $k_x$ is clearly manifested in the transmission function:
above the cutoff, the transmission decays very rapidly. 
Below the cutoff, we
see the familiar Fabry-Perot and coupled surface plasmon resonances, although
they have moved slightly; this is because the relationship between $k_x$ and
$k_z$ has been altered, as shown in figure \ref{fig:isofreq}.

Finally, we re-examine the image of a line source using the modified 
transmission functions shown in figure \ref{fig:transmission_layers}.
Figure \ref{fig:subwavelength} shows the transmitted electric field intensity,
plotted as a function of $x$, for various different layer widths. Increasing 
the width of the layers which make up the metamaterial slab (while keeping the
total slab width constant) causes the principal peak to broaden, as expected.
As the layers get thinner, the transmitted image more closely resembles the 
effective medium result.

\section{Conclusion}
We have investigated a class of anisotropic 
materials in which the one of the components of the dielectric permittivity 
has a  different sign from the others. These materials are able to 
support the propagation of modes that would normally be evanescent: they are
able to collect and transfer the near field. In addition, inside the 
anisotropic medium, light travels in a preferred direction. 

We have studied the transmission properties of a slab 
made up of such a material. 
The image of a line source consists of two lines, with an offset determined by
the ratio of the components of the permittivity; the width of the imaged lines
depends on the amount of absorption, but in principle can be much less than the
wavelength of light used.

One realization of such a material is a stack of alternating layers of metal
and dielectric. The thinner the layers, the better this metamaterial approaches
the form of the ideal anisotropic medium. We have shown how the ideal
band structure is deformed by the non-zero layer width. Using realistic material
parameters, we have also demonstrated that a stack of alternating Ag and
ZnS-SiO$_2$ layers can form an image of a line source which is much narrower 
than the wavelength of light when working at 650nm. 

The combination of subwavelength resolution with 
the fact that the position of the
image depends on the frequency of light being used suggests that this layered
system may have useful applications. For example, in conjunction with a 
super-resolution near-field optical structure (super-RENS), 
\citep{liu:near_field_images, lin:super_resolution_study}
it may allow the possibility of multiplexed recording. 

\appendix*

\section{Homogenization in layered systems}
\label{app:homogenize}
The effective medium parameters for our one-dimensional system of alternating
layers can be calculated by using the characteristic 
matrix method. The effective dielectric 
permittivity is obtained from a consideration of TM fields; TE fields give the
effective magnetic permeability. 

The geometry of the system is shown in figure \ref{fig:layered-system}, with 
the $z$-axis perpendicular to the layers. We take the plane of incidence 
to be the $xz$-plane; the symmetry of the system means that this is 
equivalent to the $yz$-plane, and the results which follow are general. 

The characteristic matrix ${\mathbf M}_j(k_x,d)$ relates the Fourier component 
of the field in the plane $z=z_0$ to that in the plane $z=z_0+d$ 
(all within medium $j$).
For TM waves in a homogeneous medium, it takes the form \citep{born:principles}
\begin{equation}
{\mathbf M}_j(k_x,d) = \begin{pmatrix}
\cos k_z^{(j)} d & \frac{ik_0\epsilon_j}{k_z^{(j)}} \sin k_z^{(j)}d \\
\frac{ik_z^{(j)}}{k_0\epsilon_j} \sin k_z^{(j)}d & \cos k_z^{(j)} d 
\end{pmatrix}
\end{equation}
where $k_z^{(j)}$ is given by the dispersion relation
\begin{equation}
k_x^2+(k_z^{(j)})^2=\epsilon k_0^2.
\label{eq:iso-disp}
\end{equation}

The matrix for a single cell of our layered system, consisting of one 
sheet of each material, is just the product of the matrices for the separate
layers:
\begin{equation}
{\mathbf M}_\mathrm{cell}(k_x, d_1, \eta) = {\mathbf M}_1 (k_x, d_1) 
{\mathbf M}_2(k_x, \eta d_1) .
\end{equation}
A stack of $n$ cells has the characteristic matrix 
${\mathbf M}_n=({\mathbf M}_\mathrm{cell})^n$. We can calculate this by 
diagonalizing ${\mathbf M}_\mathrm{cell}$; we then obtain 
\begin{equation}
{\mathbf M}_n = \frac{1}{p-q}\begin{pmatrix}
-q\lambda^n+p\lambda^{-n} & \lambda^n-\lambda^{-n} \\
-pq(\lambda^n-\lambda^{-n}) & p\lambda^n-q\lambda^{-n} \end{pmatrix}.
\end{equation}
We have introduced $\lambda$, which is one of the eigenvalues 
of ${\mathbf M}_\mathrm{cell}$; the other eigenvalue is $\lambda^{-1}$, which 
follows because $\det{\mathbf M}_\mathrm{cell}=1$. 
We have also introduced $p$ and $q$, 
which are the ratios of the components of the eigenvectors of 
${\mathbf M}_\mathrm{cell}$. 

Expanding in powers of the layer thickness allows us to relate ${\mathbf M}_n$
to the characteristic matrix for an effective medium:
\begin{align}
{\mathbf M}_n &= \begin{pmatrix}
\cos k_z^\mathrm{eff} n(1+\eta)d_1 & 
\frac{ik_0 \epsilon_x}{k_z^\mathrm{eff}}\sin k_z^\mathrm{eff} n(1+\eta)d_1 \\
\frac{ik_z^\mathrm{eff}}{k_0 \epsilon_x}\sin k_z^\mathrm{eff} n(1+\eta)d_1 & 
\cos k_z^\mathrm{eff} n(1+\eta)d_1  
\end{pmatrix}
+ \mathcal{O}\bigl[(1+\eta)d_1\bigr] \\
&= {\mathbf M}_\mathrm{eff} + \mathcal{O}\bigl[(1+\eta)d_1\bigr]
\end{align}
where the effective medium parameters are
\begin{align}
\epsilon_x &= \frac{\epsilon_1+\eta\epsilon_2}{1+\eta} \\
\frac{1}{\epsilon_z} &= \frac{1}{\epsilon_1(1+\eta)}+
\frac{\eta}{\epsilon_2(1+\eta)} .
\end{align}
These parameters appear in the dispersion relation for the effective medium,
which differs slightly from 
\eqref{eq:iso-disp} because the permittivity is now anisotropic:
\begin{equation}
\frac{k_x^2}{\epsilon_z}+\frac{(k_z^\mathrm{eff})^2}{\epsilon_x}=k_0^2.
\end{equation}

We also note here that the cell matrix ${\mathbf M}_\mathrm{cell}$ has another 
use. We can determine the true dispersion relation for the layered system 
-- without using the effective medium approximation -- by finding the 
eigenvalues and eigenvectors of this matrix. When the eigenvalue has unit 
modulus, we have found a Bloch mode; we then make the association
\begin{equation}
\lambda = e^{ik_z (1+\eta)d_1}.
\label{eq:lambda}
\end{equation}
The eigenvalue $\lambda$ depends on the frequency and on $k_x$; equation 
\eqref{eq:lambda} is therefore the dispersion relation.

\end{document}